
%
\catcode`\@=11
\font\tensmc=cmcsc10      
\def\smc{\tensmc}

\def\hcorrection#1{\advance\hoffset by #1 }
\def\vcorrection#1{\advance\voffset by #1 }
\def\wlog#1{}
\newif\iftitle@
\outer\def\title{\title@true\vglue 24\p@ plus 12\p@ minus 12\p@
   \bgroup\let\\=\cr\tabskip\centering
   \halign to \hsize\bgroup\tenbf\hfill\ignorespaces##\unskip\hfill\cr}
\def\endtitle{\cr\egroup\egroup\vglue 18\p@ plus 12\p@ minus 6\p@}
\outer\def\author{\iftitle@\vglue -18\p@ plus -12\p@ minus -6\p@\fi\vglue
    12\p@ plus 6\p@ minus 3\p@\bgroup\let\\=\cr\tabskip\centering
    \halign to \hsize\bgroup\smc\hfill\ignorespaces##\unskip\hfill\cr}
\def\endauthor{\cr\egroup\egroup\vglue 18\p@ plus 12\p@ minus 6\p@}
\outer\def\heading{\bigbreak\bgroup\let\\=\cr\tabskip\centering
    \halign to \hsize\bgroup\smc\hfill\ignorespaces##\unskip\hfill\cr}
\def\endheading{\cr\egroup\egroup\nobreak\medskip}

\outer\def\endproclaim{\par\ifdim\lastskip<\medskipamount\removelastskip
  \penalty 55 \fi\medskip\rm}
\outer\def\demo#1{\par\ifdim\lastskip<\smallskipamount\removelastskip
    \smallskip\fi\noindent{\smc\ignorespaces#1\unskip:\enspace}\rm
      \ignorespaces}

\newcount\footmarkcount@
\footmarkcount@=1
\def\makefootnote@#1#2{\insert\footins{\interlinepenalty=100
  \splittopskip=\ht\strutbox \splitmaxdepth=\dp\strutbox
  \floatingpenalty=\@MM
  \leftskip=\z@\rightskip=\z@\spaceskip=\z@\xspaceskip=\z@
  \noindent{#1}\footstrut\rm\ignorespaces #2\strut}}
\def\footnote{\let\@sf=\empty\ifhmode\edef\@sf{\spacefactor
   =\the\spacefactor}\/\fi\futurelet\next\footnote@}
\def\footnote@{\ifx"\next\let\next\footnote@@\else
    \let\next\footnote@@@\fi\next}
\def\footnote@@"#1"#2{#1\@sf\relax\makefootnote@{#1}{#2}}
\def\footnote@@@#1{$^{\number\footmarkcount@}$\makefootnote@
   {$^{\number\footmarkcount@}$}{#1}\global\advance\footmarkcount@ by 1 }

\hyphenation{man-u-script man-u-scripts ap-pen-dix ap-pen-di-ces}
\hyphenation{data-base data-bases}
\ifx\amstexloaded@\relax\catcode`\@=13
  \endinput\else\let\amstexloaded@=\relax\fi
\newlinechar=`\^^J
\def\eat@#1{}
\def\Space@.{\futurelet\Space@\relax}
\Space@. %
\newhelp\athelp@
{Only certain combinations beginning with @ make sense to me.^^J
Perhaps you wanted \string\@\space for a printed @?^^J
I've ignored the character or group after @.}
\def\futureletnextat@{\futurelet\next\at@}
{\catcode`\@=\active
\lccode`\Z=`\@ \lowercase
{\gdef@{\expandafter\csname futureletnextatZ\endcsname}
\expandafter\gdef\csname atZ\endcsname
   {\ifcat\noexpand\next a\def\next{\csname atZZ\endcsname}\else
   \ifcat\noexpand\next0\def\next{\csname atZZ\endcsname}\else
    \def\next{\csname atZZZ\endcsname}\fi\fi\next}
\expandafter\gdef\csname atZZ\endcsname#1{\expandafter
   \ifx\csname #1Zat\endcsname\relax\def\next
     {\errhelp\expandafter=\csname athelpZ\endcsname
      \errmessage{Invalid use of \string@}}\else
       \def\next{\csname #1Zat\endcsname}\fi\next}
\expandafter\gdef\csname atZZZ\endcsname#1{\errhelp
    \expandafter=\csname athelpZ\endcsname
      \errmessage{Invalid use of \string@}}}}
\def\atdef@#1{\expandafter\def\csname #1@at\endcsname}
\newhelp\defahelp@{If you typed \string\define\space cs instead of
\string\define\string\cs\space^^J
I've substituted an inaccessible control sequence so that your^^J
definition will be completed without mixing me up too badly.^^J
If you typed \string\define{\string\cs} the inaccessible control sequence^^J
was defined to be \string\cs, and the rest of your^^J
definition appears as input.}
\newhelp\defbhelp@{I've ignored your definition, because it might^^J
conflict with other uses that are important to me.}
\def\define{\futurelet\next\define@}
\def\define@{\ifcat\noexpand\next\relax
  \def\next{\define@@}%
  \else\errhelp=\defahelp@
  \errmessage{\string\define\space must be followed by a control
     sequence}\def\next{\def\garbage@}\fi\next}
\def\undefined@{}
\def\preloaded@{}
\def\define@@#1{\ifx#1\relax\errhelp=\defbhelp@
   \errmessage{\string#1\space is already defined}\def\next{\def\garbage@}%
   \else\expandafter\ifx\csname\expandafter\eat@\string
         #1@\endcsname\undefined@\errhelp=\defbhelp@
   \errmessage{\string#1\space can't be defined}\def\next{\def\garbage@}%
   \else\expandafter\ifx\csname\expandafter\eat@\string#1\endcsname\relax
     \def\next{\def#1}\else\errhelp=\defbhelp@
     \errmessage{\string#1\space is already defined}\def\next{\def\garbage@}%
      \fi\fi\fi\next}
\def\famzero{\fam\z@}

\def\lim{\mathop{\famzero lim}}

\def\textfont@#1#2{\def#1{\relax\ifmmode
    \errmessage{Use \string#1\space only in text}\else#2\fi}}
\textfont@\rm\tenrm
\textfont@\it\tenit
\textfont@\sl\tensl
\textfont@\bf\tenbf
\textfont@\smc\tensmc
\let\ic@=\/
\def\/{\unskip\ic@}
\def\textfonti{\the\textfont1 }
\def\t#1#2{{\edef\next{\the\font}\textfonti\accent"7F \next#1#2}}
\let\B=\=
\let\D=\.
\def~{\unskip\nobreak\ \ignorespaces}
{\catcode`\@=\active
\gdef\@{\char'100 }}
\atdef@-{\leavevmode\futurelet\next\athyph@}
\def\athyph@{\ifx\next-\let\next=\athyph@@
  \else\let\next=\athyph@@@\fi\next}
\def\athyph@@@{\hbox{-}}
\def\athyph@@#1{\futurelet\next\athyph@@@@}
\def\athyph@@@@{\if\next-\def\next##1{\hbox{---}}\else
    \def\next{\hbox{--}}\fi\next}
\def\.{.\spacefactor=\@m}
\atdef@.{\null.}
\atdef@,{\null,}
\atdef@;{\null;}
\atdef@:{\null:}
\atdef@?{\null?}
\atdef@!{\null!}
\def\srdr@{\thinspace}
\def\drsr@{\kern.02778em}
\def\sldl@{\kern.02778em}
\def\dlsl@{\thinspace}
\atdef@"{\unskip\futurelet\next\atqq@}
\def\atqq@{\ifx\next\Space@\def\next. {\atqq@@}\else
         \def\next.{\atqq@@}\fi\next.}
\def\atqq@@{\futurelet\next\atqq@@@}
\def\atqq@@@{\ifx\next`\def\next`{\atqql@}\else\def\next'{\atqqr@}\fi\next}
\def\atqql@{\futurelet\next\atqql@@}
\def\atqql@@{\ifx\next`\def\next`{\sldl@``}\else\def\next{\dlsl@`}\fi\next}
\def\atqqr@{\futurelet\next\atqqr@@}
\def\atqqr@@{\ifx\next'\def\next'{\srdr@''}\else\def\next{\drsr@'}\fi\next}

\def\textfontii{\the\textfont2 }
\def\{{\relax\ifmmode\lbrace\else
    {\textfontii f}\spacefactor=\@m\fi}
\def\}{\relax\ifmmode\rbrace\else
    \let\@sf=\empty\ifhmode\edef\@sf{\spacefactor=\the\spacefactor}\fi
      {\textfontii g}\@sf\relax\fi}
\def\nonhmodeerr@#1{\errmessage
     {\string#1\space allowed only within text}}
\def\linebreak{\relax\ifhmode\unskip\break\else
    \nonhmodeerr@\linebreak\fi}
\def\allowlinebreak{\relax
   \ifhmode\allowbreak\else\nonhmodeerr@\allowlinebreak\fi}
\newskip\saveskip@
\def\nolinebreak{\relax\ifhmode\saveskip@=\lastskip\unskip
  \nobreak\ifdim\saveskip@>\z@\hskip\saveskip@\fi
   \else\nonhmodeerr@\nolinebreak\fi}
\def\newline{\relax\ifhmode\null\hfil\break
    \else\nonhmodeerr@\newline\fi}
\def\nonmathaerr@#1{\errmessage
     {\string#1\space is not allowed in display math mode}}
\def\nonmathberr@#1{\errmessage{\string#1\space is allowed only in math mode}}
\def\mathbreak{\relax\ifmmode\ifinner\break\else
   \nonmathaerr@\mathbreak\fi\else\nonmathberr@\mathbreak\fi}
\def\nomathbreak{\relax\ifmmode\ifinner\nobreak\else
    \nonmathaerr@\nomathbreak\fi\else\nonmathberr@\nomathbreak\fi}
\def\allowmathbreak{\relax\ifmmode\ifinner\allowbreak\else
     \nonmathaerr@\allowmathbreak\fi\else\nonmathberr@\allowmathbreak\fi}
\def\pagebreak{\relax\ifmmode
   \ifinner\errmessage{\string\pagebreak\space
     not allowed in non-display math mode}\else\postdisplaypenalty-\@M\fi
   \else\ifvmode\penalty-\@M\else\edef\spacefactor@
       {\spacefactor=\the\spacefactor}\vadjust{\penalty-\@M}\spacefactor@
        \relax\fi\fi}
\def\nopagebreak{\relax\ifmmode
     \ifinner\errmessage{\string\nopagebreak\space
    not allowed in non-display math mode}\else\postdisplaypenalty\@M\fi
    \else\ifvmode\nobreak\else\edef\spacefactor@
        {\spacefactor=\the\spacefactor}\vadjust{\penalty\@M}\spacefactor@
         \relax\fi\fi}
\def\newpage{\relax\ifvmode\vfill\penalty-\@M\else\nonvmodeerr@\newpage\fi}
\def\nonvmodeerr@#1{\errmessage
    {\string#1\space is allowed only between paragraphs}}
\def\smallpagebreak{\relax\ifvmode\smallbreak
      \else\nonvmodeerr@\smallpagebreak\fi}
\def\medpagebreak{\relax\ifvmode\medbreak
       \else\nonvmodeerr@\medpagebreak\fi}
\def\bigpagebreak{\relax\ifvmode\bigbreak
      \else\nonvmodeerr@\bigpagebreak\fi}
\newdimen\captionwidth@
\captionwidth@=\hsize
\advance\captionwidth@ by -1.5in
\def\caption#1{}
\def\topspace#1{\gdef\thespace@{#1}\ifvmode\def\next
    {\futurelet\next\topspace@}\else\def\next{\nonvmodeerr@\topspace}\fi\next}
\def\topspace@{\ifx\next\Space@\def\next. {\futurelet\next\topspace@@}\else
     \def\next.{\futurelet\next\topspace@@}\fi\next.}
\def\topspace@@{\ifx\next\caption\let\next\topspace@@@\else
    \let\next\topspace@@@@\fi\next}
 \def\topspace@@@@{\topinsert\vbox to
       \thespace@{}\endinsert}
\def\topspace@@@\caption#1{\topinsert\vbox to
    \thespace@{}\nobreak
      \smallskip
    \setbox\z@=\hbox{\noindent\ignorespaces#1\unskip}%
   \ifdim\wd\z@>\captionwidth@
   \centerline{\vbox{\hsize=\captionwidth@\noindent\ignorespaces#1\unskip}}%
   \else\centerline{\box\z@}\fi\endinsert}
\def\midspace#1{\gdef\thespace@{#1}\ifvmode\def\next
    {\futurelet\next\midspace@}\else\def\next{\nonvmodeerr@\midspace}\fi\next}
\def\midspace@{\ifx\next\Space@\def\next. {\futurelet\next\midspace@@}\else
     \def\next.{\futurelet\next\midspace@@}\fi\next.}
\def\midspace@@{\ifx\next\caption\let\next\midspace@@@\else
    \let\next\midspace@@@@\fi\next}
 \def\midspace@@@@{\midinsert\vbox to
       \thespace@{}\endinsert}
\def\midspace@@@\caption#1{\midinsert\vbox to
    \thespace@{}\nobreak
      \smallskip
      \setbox\z@=\hbox{\noindent\ignorespaces#1\unskip}%
      \ifdim\wd\z@>\captionwidth@
    \centerline{\vbox{\hsize=\captionwidth@\noindent\ignorespaces#1\unskip}}%
    \else\centerline{\box\z@}\fi\endinsert}
\mathchardef\prime@="0230
\def\prime{{{}\prime@{}}}
\def\prim@s{\prime@\futurelet\next\pr@m@s}

\def\,{\relax\ifmmode\mskip\thinmuskip\else\thinspace\fi}
\def\!{\relax\ifmmode\mskip-\thinmuskip\else\negthinspace\fi}
\def\frac#1#2{{#1\over#2}}

\def\:{\nobreak\hskip.1111em{:}\hskip.3333em plus .0555em\relax}
\def\intic@{\mathchoice{\hskip5\p@}{\hskip4\p@}{\hskip4\p@}{\hskip4\p@}}
\def\negintic@
 {\mathchoice{\hskip-5\p@}{\hskip-4\p@}{\hskip-4\p@}{\hskip-4\p@}}
\def\intkern@{\mathchoice{\!\!\!}{\!\!}{\!\!}{\!\!}}
\def\intdots@{\mathchoice{\cdots}{{\cdotp}\mkern1.5mu
    {\cdotp}\mkern1.5mu{\cdotp}}{{\cdotp}\mkern1mu{\cdotp}\mkern1mu
      {\cdotp}}{{\cdotp}\mkern1mu{\cdotp}\mkern1mu{\cdotp}}}
\newcount\intno@
\def\iint{\intno@=\tw@\futurelet\next\ints@}
\def\iiint{\intno@=\thr@@\futurelet\next\ints@}
\def\iiiint{\intno@=4 \futurelet\next\ints@}
\def\idotsint{\intno@=\z@\futurelet\next\ints@}
\def\ints@{\findlimits@\ints@@}
\newif\iflimtoken@
\newif\iflimits@
\def\findlimits@{\limtoken@false\limits@false\ifx\next\limits
 \limtoken@true\limits@true\else\ifx\next\nolimits\limtoken@true\limits@false
    \fi\fi}
\def\multintlimits@{\intop\ifnum\intno@=\z@\intdots@
  \else\intkern@\fi
    \ifnum\intno@>\tw@\intop\intkern@\fi
     \ifnum\intno@>\thr@@\intop\intkern@\fi\intop}
\def\multint@{\int\ifnum\intno@=\z@\intdots@\else\intkern@\fi
   \ifnum\intno@>\tw@\int\intkern@\fi
    \ifnum\intno@>\thr@@\int\intkern@\fi\int}
\def\ints@@{\iflimtoken@\def\ints@@@{\iflimits@
   \negintic@\mathop{\intic@\multintlimits@}\limits\else
    \multint@\nolimits\fi\eat@}\else
     \def\ints@@@{\multint@\nolimits}\fi\ints@@@}
\def\Sb{_\bgroup\vspace@
        \baselineskip=\fontdimen10 \scriptfont\tw@
        \advance\baselineskip by \fontdimen12 \scriptfont\tw@
        \lineskip=\thr@@\fontdimen8 \scriptfont\thr@@
        \lineskiplimit=\thr@@\fontdimen8 \scriptfont\thr@@
        \Let@\vbox\bgroup\halign\bgroup \hfil$\scriptstyle
            {##}$\hfil\cr}
\def\endSb{\crcr\egroup\egroup\egroup}
\def\Sp{^\bgroup\vspace@
        \baselineskip=\fontdimen10 \scriptfont\tw@
        \advance\baselineskip by \fontdimen12 \scriptfont\tw@
        \lineskip=\thr@@\fontdimen8 \scriptfont\thr@@
        \lineskiplimit=\thr@@\fontdimen8 \scriptfont\thr@@
        \Let@\vbox\bgroup\halign\bgroup \hfil$\scriptstyle
            {##}$\hfil\cr}
\def\endSp{\crcr\egroup\egroup\egroup}
\def\Let@{\relax\iffalse{\fi\let\\=\cr\iffalse}\fi}
\def\vspace@{\def\vspace##1{\noalign{\vskip##1 }}}
\def\aligned{\,\vcenter\bgroup\vspace@\Let@\openup\jot\m@th\ialign
  \bgroup \strut\hfil$\displaystyle{##}$&$\displaystyle{{}##}$\hfil\crcr}
\def\endaligned{\crcr\egroup\egroup}
\def\matrix{\,\vcenter\bgroup\Let@\vspace@
    \normalbaselines
  \m@th\ialign\bgroup\hfil$##$\hfil&&\quad\hfil$##$\hfil\crcr
    \mathstrut\crcr\noalign{\kern-\baselineskip}}
\def\endmatrix{\crcr\mathstrut\crcr\noalign{\kern-\baselineskip}\egroup
                \egroup\,}
\newtoks\hashtoks@
\hashtoks@={#}
\def\format{\crcr\egroup\iffalse{\fi\ifnum`}=0 \fi\format@}
\def\format@#1\\{\def\preamble@{#1}%
  \def\c{\hfil$\the\hashtoks@$\hfil}%
  \def\r{\hfil$\the\hashtoks@$}%
  \def\l{$\the\hashtoks@$\hfil}%
  \setbox\z@=\hbox{\xdef\Preamble@{\preamble@}}\ifnum`{=0 \fi\iffalse}\fi
   \ialign\bgroup\span\Preamble@\crcr}

\def\cases{\left\{\,\vcenter\bgroup\vspace@
     \normalbaselines\openup\jot\m@th
       \Let@\ialign\bgroup$##$\hfil&\quad$##$\hfil\crcr
      \mathstrut\crcr\noalign{\kern-\baselineskip}}

\newif\iftagsleft@
\tagsleft@true
\def\TagsOnRight{\global\tagsleft@false}
\def\tag#1$${\iftagsleft@\leqno\else\eqno\fi
 \hbox{\def\pagebreak{\global\postdisplaypenalty-\@M}%
 \def\nopagebreak{\global\postdisplaypenalty\@M}\rm(#1\unskip)}%
  $$\postdisplaypenalty\z@\ignorespaces}
\interdisplaylinepenalty=\@M
\def\allowdisplaybreak@{\def\allowdisplaybreak{\noalign{\allowbreak}}}
\def\displaybreak@{\def\displaybreak{\noalign{\break}}}
\def\align#1\endalign{\def\tag{&}\vspace@\allowdisplaybreak@\displaybreak@
  \iftagsleft@\lalign@#1\endalign\else
   \ralign@#1\endalign\fi}
\def\ralign@#1\endalign{\displ@y\Let@\tabskip\centering\halign to\displaywidth
     {\hfil$\displaystyle{##}$\tabskip=\z@&$\displaystyle{{}##}$\hfil
       \tabskip=\centering&\llap{\hbox{(\rm##\unskip)}}\tabskip\z@\crcr
             #1\crcr}}
\def\lalign@
 #1\endalign{\displ@y\Let@\tabskip\centering\halign to \displaywidth
   {\hfil$\displaystyle{##}$\tabskip=\z@&$\displaystyle{{}##}$\hfil
   \tabskip=\centering&\kern-\displaywidth
        \rlap{\hbox{(\rm##\unskip)}}\tabskip=\displaywidth\crcr
               #1\crcr}}
\def\overrightarrow{\mathpalette\overrightarrow@}
\def\overrightarrow@#1#2{\vbox{\ialign{$##$\cr
    #1{-}\mkern-6mu\cleaders\hbox{$#1\mkern-2mu{-}\mkern-2mu$}\hfill
     \mkern-6mu{\to}\cr
     \noalign{\kern -1\p@\nointerlineskip}
     \hfil#1#2\hfil\cr}}}
\def\overleftarrow{\mathpalette\overleftarrow@}
\def\overleftarrow@#1#2{\vbox{\ialign{$##$\cr
     #1{\leftarrow}\mkern-6mu\cleaders\hbox{$#1\mkern-2mu{-}\mkern-2mu$}\hfill
      \mkern-6mu{-}\cr
     \noalign{\kern -1\p@\nointerlineskip}
     \hfil#1#2\hfil\cr}}}
\def\overleftrightarrow{\mathpalette\overleftrightarrow@}
\def\overleftrightarrow@#1#2{\vbox{\ialign{$##$\cr
     #1{\leftarrow}\mkern-6mu\cleaders\hbox{$#1\mkern-2mu{-}\mkern-2mu$}\hfill
       \mkern-6mu{\to}\cr
    \noalign{\kern -1\p@\nointerlineskip}
      \hfil#1#2\hfil\cr}}}
\def\underrightarrow{\mathpalette\underrightarrow@}
\def\underrightarrow@#1#2{\vtop{\ialign{$##$\cr
    \hfil#1#2\hfil\cr
     \noalign{\kern -1\p@\nointerlineskip}
    #1{-}\mkern-6mu\cleaders\hbox{$#1\mkern-2mu{-}\mkern-2mu$}\hfill
     \mkern-6mu{\to}\cr}}}
\def\underleftarrow{\mathpalette\underleftarrow@}
\def\underleftarrow@#1#2{\vtop{\ialign{$##$\cr
     \hfil#1#2\hfil\cr
     \noalign{\kern -1\p@\nointerlineskip}
     #1{\leftarrow}\mkern-6mu\cleaders\hbox{$#1\mkern-2mu{-}\mkern-2mu$}\hfill
      \mkern-6mu{-}\cr}}}
\def\underleftrightarrow{\mathpalette\underleftrightarrow@}
\def\underleftrightarrow@#1#2{\vtop{\ialign{$##$\cr
      \hfil#1#2\hfil\cr
    \noalign{\kern -1\p@\nointerlineskip}
     #1{\leftarrow}\mkern-6mu\cleaders\hbox{$#1\mkern-2mu{-}\mkern-2mu$}\hfill
       \mkern-6mu{\to}\cr}}}
\def\sqrt#1{\radical"270370 {#1}}
\def\dots{\relax\ifmmode\let\next=\ldots\else\let\next=\tdots@\fi\next}
\def\tdots@{\unskip\ \tdots@@}
\def\tdots@@{\futurelet\next\tdots@@@}
\def\tdots@@@{$\mathinner{\ldotp\ldotp\ldotp}\,
   \ifx\next,$\else
   \ifx\next.\,$\else
   \ifx\next;\,$\else
   \ifx\next:\,$\else
   \ifx\next?\,$\else
   \ifx\next!\,$\else
   $ \fi\fi\fi\fi\fi\fi}
\def\text{\relax\ifmmode\let\next=\text@\else\let\next=\text@@\fi\next}
\def\text@@#1{\hbox{#1}}
\def\text@#1{\mathchoice
 {\hbox{\everymath{\displaystyle}\def\textfonti{\the\textfont1 }%
    \def\textfontii{\the\textfont2 }\textdef@@ T#1}}
 {\hbox{\everymath{\textstyle}\def\textfonti{\the\textfont1 }%
    \def\textfontii{\the\textfont2 }\textdef@@ T#1}}
 {\hbox{\everymath{\scriptstyle}\def\textfonti{\the\scriptfont1 }%
   \def\textfontii{\the\scriptfont2 }\textdef@@ S\rm#1}}
 {\hbox{\everymath{\scriptscriptstyle}\def\textfonti{\the\scriptscriptfont1 }%
   \def\textfontii{\the\scriptscriptfont2 }\textdef@@ s\rm#1}}}
\def\textdef@@#1{\textdef@#1\rm \textdef@#1\bf
   \textdef@#1\sl \textdef@#1\it}

\def\textdef@#1#2{\def\next{\csname\expandafter\eat@\string#2fam\endcsname}%
\if S#1\edef#2{\the\scriptfont\next\relax}%
 \else\if s#1\edef#2{\the\scriptscriptfont\next\relax}%
 \else\edef#2{\the\textfont\next\relax}\fi\fi}
\scriptfont\itfam=\tenit \scriptscriptfont\itfam=\tenit
\scriptfont\slfam=\tensl \scriptscriptfont\slfam=\tensl
\mathcode`\0="0030
\mathcode`\1="0031
\mathcode`\2="0032
\mathcode`\3="0033
\mathcode`\4="0034
\mathcode`\5="0035
\mathcode`\6="0036
\mathcode`\7="0037
\mathcode`\8="0038
\mathcode`\9="0039
\def\Cal{\relax\ifmmode\let\next=\Cal@\else
     \def\next{\errmessage{Use \string\Cal\space only in math mode}}\fi\next}
\def\Cal@#1{{\fam2 #1}}
\def\bold{\relax\ifmmode\let\next=\bold@\else
   \def\next{\errmessage{Use \string\bold\space only in math
      mode}}\fi\next}\def\bold@#1{{\fam\bffam #1}}
\mathchardef\Gamma="0000
\mathchardef\Delta="0001
\mathchardef\Theta="0002
\mathchardef\Lambda="0003
\mathchardef\Xi="0004
\mathchardef\Pi="0005
\mathchardef\Sigma="0006
\mathchardef\Upsilon="0007
\mathchardef\Phi="0008
\mathchardef\Psi="0009
\mathchardef\Omega="000A
\mathchardef\varGamma="0100
\mathchardef\varDelta="0101
\mathchardef\varTheta="0102
\mathchardef\varLambda="0103
\mathchardef\varXi="0104
\mathchardef\varPi="0105
\mathchardef\varSigma="0106
\mathchardef\varUpsilon="0107
\mathchardef\varPhi="0108
\mathchardef\varPsi="0109
\mathchardef\varOmega="010A
\font\dummyft@=dummy
\fontdimen1 \dummyft@=\z@
\fontdimen2 \dummyft@=\z@
\fontdimen3 \dummyft@=\z@
\fontdimen4 \dummyft@=\z@
\fontdimen5 \dummyft@=\z@
\fontdimen6 \dummyft@=\z@
\fontdimen7 \dummyft@=\z@
\fontdimen8 \dummyft@=\z@
\fontdimen9 \dummyft@=\z@
\fontdimen10 \dummyft@=\z@
\fontdimen11 \dummyft@=\z@
\fontdimen12 \dummyft@=\z@
\fontdimen13 \dummyft@=\z@
\fontdimen14 \dummyft@=\z@
\fontdimen15 \dummyft@=\z@
\fontdimen16 \dummyft@=\z@
\fontdimen17 \dummyft@=\z@
\fontdimen18 \dummyft@=\z@
\fontdimen19 \dummyft@=\z@
\fontdimen20 \dummyft@=\z@
\fontdimen21 \dummyft@=\z@
\fontdimen22 \dummyft@=\z@
\def\fontlist@{\\{\tenrm}\\{\sevenrm}\\{\fiverm}\\{\teni}\\{\seveni}%
 \\{\fivei}\\{\tensy}\\{\sevensy}\\{\fivesy}\\{\tenex}\\{\tenbf}\\{\sevenbf}%
 \\{\fivebf}\\{\tensl}\\{\tenit}\\{\tensmc}}
\def\dodummy@{{\def\\##1{\global\let##1=\dummyft@}\fontlist@}}
\newif\ifsyntax@
\newcount\countxviii@
\def\newtoks@{\alloc@5\toks\toksdef\@cclvi}
\def\nopages@{\output={\setbox\z@=\box\@cclv \deadcycles=\z@}\newtoks@\output}
\def\syntax{\syntax@true\dodummy@\countxviii@=\count18
\loop \ifnum\countxviii@ > \z@ \textfont\countxviii@=\dummyft@
   \scriptfont\countxviii@=\dummyft@ \scriptscriptfont\countxviii@=\dummyft@
     \advance\countxviii@ by-\@ne\repeat
\dummyft@\tracinglostchars=\z@
  \nopages@\frenchspacing\hbadness=\@M}
\def\magstep#1{\ifcase#1 1000\or
 1200\or 1440\or 1728\or 2074\or 2488\or
 \errmessage{\string\magstep\space only works up to 5}\fi\relax}
{\lccode`\2=`\p \lccode`\3=`\t
 \lowercase{\gdef\tru@#123{#1truept}}}

\def\scaletype#1{\mag=#1\relax
 \hsize=\expandafter\tru@\the\hsize
 \vsize=\expandafter\tru@\the\vsize
 \dimen\footins=\expandafter\tru@\the\dimen\footins}

\def\scalefont#1#2\andcallit#3{\edef\font@{\the\font}#1\font#3=
  \fontname\font\space scaled #2\relax\font@}
\def\Mag@#1#2{\ifdim#1<1pt\multiply#1 #2\relax\divide#1 1000 \else
  \ifdim#1<10pt\divide#1 10 \multiply#1 #2\relax\divide#1 100\else
  \divide#1 100 \multiply#1 #2\relax\divide#1 10 \fi\fi}
\def\scalelinespacing#1{\Mag@\baselineskip{#1}\Mag@\lineskip{#1}%
  \Mag@\lineskiplimit{#1}}
\def\wlog#1{\immediate\write-1{#1}}
\catcode`\@=\active
%
\let\vamatrix\matrix
\def\matrix#1{\vamatrix #1 \endmatrix}
\let\vabf\bf
\let\varm\rm
\def\bf#1{\ifmmode\bold#1\else\vabf#1\fi}
\def\rm#1{\ifmmode\famzero#1\else\varm#1\fi}
\font\tenbf=cmbx10
\font\tenrm=cmr10
\font\tenit=cmti10

\font\eightrm=cmr8
\font\eightit=cmti8
\font\sevenrm=cmr7
\def\sectiontitle#1\par{\vskip0pt plus.1\vsize\penalty-250
 \vskip0pt plus-.1\vsize\bigskip\vskip\parskip
 \message{#1}\leftline{\tenbf#1}\nobreak\vglue 5pt}
\def\qed{\hbox{${\vcenter{\vbox{
    \hrule height 0.4pt\hbox{\vrule width 0.4pt height 6pt
    \kern5pt\vrule width 0.4pt}\hrule height 0.4pt}}}$}}
\TagsOnRight
\hsize=5.0truein
\vsize=7.8truein
\parindent=15pt
\nopagenumbers
\baselineskip=10pt
\rightline{Revised Version of RIMS-881}
\vglue 5pc
\baselineskip=13pt
\headline{\ifnum\pageno=1\hfil\else%
{\ifodd\pageno\rightheadline \else \leftheadline\fi}\fi}
\def\rightheadline{\hfil\eightit
Hamiltonian Reduction of Super
Kac-Moody Algebras
\quad\eightrm\folio}
\def\leftheadline{\eightrm\folio\quad
\eightit
Kazuhiro Kimura
\hfil}
\voffset=2\baselineskip
\centerline{\tenbf
HAMILTONIAN REDUCTION OF SUPER OSP(1,2)}
\centerline{\tenbf AND SL(2,1)
KAC-MOODY ALGEBRAS\footnote"$^{*}$"
{To appear in Proceedings of the RIMS Research Project 91
 on Infinite Analysis}}
\vskip 1 truecm
\centerline
{\eightrm
KAZUHIRO KIMURA
}

\centerline{\eightit
}
\baselineskip=10pt
\centerline{\eightit
Research Institute for Mathematical Sciences
}
\centerline{\eightit Kyoto University, Kyoto 606, Japan
}
\vglue 20pt
\vglue 16pt
\vglue 20pt
\centerline{\eightrm ABSTRACT}
{\rightskip=1.5pc
\leftskip=1.5pc
\eightrm\parindent=1pc
We present the Wakimoto construction of the super  OSp(1,2)
and SL(2,1) Kac-Moody algebras and the free field representation
of the corresponding WZW models. After imposing suitable constraints,
we can lead the Feigin-Fuchs representation of Virasoro algebras
and coadjoint actions ofthe N=1 and N=2 conformal symmetries.
This formulation corresponds to a supercovariant extension
of the Drinfeld and Sokolov Hamiltonian reduction.
\vglue12pt}
\baselineskip=13pt
%

\overfullrule=0pt
\def\bg{\bf}
\input mssymb.tex

\def\references{{\leftline{\bg References}} \vskip 3mm}

\def\chapter#1{\centerline{\bg#1} \vskip 5mm}
\def\section#1{\leftline{\bg#1} \vskip 5mm}
\def\endpage{\vfill \eject}

\def\coderm{\hat D_-}
\def\coder+{\hat D_+}
\def\curm{g^{-1}\hat D_-g}
\def\cur+{g^{-1}\hat D_+g}
\def\curr{ Dgg^{-1}}
\def\curn{g^{-1}D_-g}
\def\psid{\Psi^\dagger}
\def\coor{(x_+,x_-,\theta_+,\theta_-)}
\def\difo{{\cal L}}

\def\refa
{\item{[1]}
V.G.Drinfeld and V.V.Sokolov, Sov. J. Math. {\bf 30} (1985) 1975.}

\def\refb
{\item{[2]}
M.Bershadsky and H.Ooguri, Commun. Math. Phys. {\bf 126} (1989) 49.}

\def\refc
{\item{[3]}
L.D.Faddeev, Commun. Math. Phys. {\bf 131} (1990) 131.}

\def\refd
{\item{[4]}
A.Alekseev and S.Shatashivili, Commun. Math. Phys. {\bf 133} (1990) 353.}

\def\refe
{\item{[5]}
T.Inami and H. Kanno, Commun. Math. Phys. {\bf 136} (1991) 605.}

\def\reff
{\item{[6]}
P.Mathieu, Phys. Lett. {\bf B218} (1989) 185.}

\def\refg
{\item{[7]}
M.Chaichian and P. P.Kulish, Phys. Lett. {\bf B183} (1987) 169.}

\def\refh
{\item{[8]}
H.Nohara and K.Mohri, Nucl. Phys. {\bf B349} (1991) 253.}

\def\refi
{\item{[9]}
S.Komata, K.Mohri and H.Nohara, Nucl. Phys. {\bf B359} (1991) 168.}

\def\refj
{\item{[10]}
M.Wakimoto, Commun. Math. Phys. {\bf 104} (1986) 605.}

\def\refk
{\item{[11]}
M.Bershadsky and H.Ooguri, Phys. Lett. {\bf B229} (1989) 374.}

\def\refl
{\item{[12]}
K.Ito, Phys. Lett. {\bf B259} (1991) 73; K.Ito, ``N=2 superconformal
$\rm CP_n$ models'',Kyoto preprint, YITP/K-934.}

\def\refm
{\item{[13]}
T.Inami and K.-I. Izawa, Phys. Lett. {\bf B255} (1991) 521.}

\def\refmm
{\item{}
Izawa K.-I., {\tenit``An Interplay between Super Structures of
Base and Target Spaces''}, Kyoto preprint, KUNS 1072 HE(TH)91/06}

\def\refn
{\item{[14]}
T.Kuramoto, {\tenit``Quantum Hamiltonian Reduction of Super Kac-Moody
Algebra''}, Queen Mary College preprint, QMW-PH-90-20.}

\def\refo
{\item{[15]}
K.Kimura, Phys. Lett. {\bf B252} (1990) 370.}

\def\refp
{\item{[16]}
S.Aoyama, Phys. Lett. {\bf B288} (1989) 355.}

\def\refq
{\item{[17]}
G.W.Delius et al. Int. Jour. Mod. Phys. {\bf A5} (1990) 3943.}

\def\refr
{\item{[18]}
M.A.Olshanetsky, Commun. Math. Phys. {\bf 88} (1983) 63.}
{\item{}

\def\refrr
{\item{}
D.A.Leites, M.V.Saveliev and V.V.Serganova, in Group Theoretical methods
in Physics, Vol.1, eds. M.A.Markov et al. (VNU Science Press, Haarlem,1986).}

\def\refs
{\item{[19]}
E.Witten, {\tenit``The N Matrix Model and Gauged WZW Models''},
preprint
IASSNS-HEP-91-26.}

\overfullrule=0pt

\sectiontitle{ 1. Introduction}

\noindent

There have recently been remarkable developments in the study of
two dimensional systems,
for example, conformal field theories, integrable models,
two dimensional gravity and topological field theories.
These theories are closely related to each other.
In particular, there are various connections between
Wess-Zumino-Witten(WZW) models and these theories.
WZW models have a symmetry associated with the Kac-Moody(KM)
algebras and provide us with a useful method for the
characterization of the chiral algebraic structure of rational
conformal field theories.
\par
It is well-known that there are two methods of constructing
extended conformal symmetries, for example,  super Virasoro
algebras and W algebras.
One is the coset construction and the other is the Hamiltonian
reduction of KM algebras[1]. Both of the methods can be studied by
realizing them as gauged WZW models, which give a Lagrangian
approach and a geometric interpretation of
symmetries[2][3][4].
Drinfeld and Sokolov first succeeded in giving the
relation among the W algebra, the generalized KdV equations
and the Toda field theories by means of the Hamiltonian reduction.
There are several attempts extending the Hamiltonian reduction to
supersymmetric ones[5][6][7].
 Komata et al. considered
the series of Lie superalgebra which lead to the supersymmetric
Toda field theories[8][9].
They investigated two series of the extended
superconformal algebras associated with the affine Lie
 superalgebras $\rm SL(n+1,n)(n\geq 1)$ and
$\rm OSp(2n\pm1,2n)(n\geq 1)$
and gave the Feigin-Fuchs construction
of these algebras.
On the other hand, the free field realization
is also a powerful method to investigate
two-dimensional systems systematically. The Feigin-Fuchs construction
provides us with means to describe extended conformal algebras in the Fock
space of free fields and to calculate characters and correlation
functions by means of BRST cohomology. This realization can be
applied to KM algebras with an arbitrary central charge, called
Wakimoto construction[10]. The currents of algebras
are expressed
by free bosons corresponding to the Cartan part and ghosts for roots.
\par
It is transparent to perform the Hamiltonian reduction based on
the free field realization.
Bershadsky and Ooguri first studied the Hamiltonian reduction of
$\rm Osp(1,2)$ KM algebra in terms of free fields[11].
Ito[12] has extended their method to Lie superalgebras. He showed
that the quantum Hamiltonian reduction of the affine
Lie superalgebras $\rm SL(n+1,n)(n\geq 1)$ has a W algebra
 structure with the N=2 superconformal symmetry and leads
to the N=2 $\rm CP_n=SU(n+1)/SU(n)\times U(1)$ coset model
constructed by Kazama and Suzuki.
Inami and Izawa proposed the Hamiltonian reduction of WZW models based on
Lie supergroups with a fermionic simple root system[13].
They obtained actions of the super-Toda theories
and also geometric actions of the superconformal groups.
In those papers,however it was inevitably necessary
to introduce extra fermions by hand.
Kuramoto pointed out that the Hamiltonian reduction of super $\rm
OSp(1,2)$KM algebras does not
need fermionic auxiliary fields because the superpartners of the bosonic
currents play the role of them[14].
Nevertheless, his formulation is not supercovariant and may not probably be
adequate to general Lie superalgebras.
\par
In this paper we shall present explicitly the Wakimoto construction
of the super $\rm OSp(1,2)$ and $\rm SL(2,1)$[14] KM algebras,
in other words, the free boson representation
of the super WZW models and perform the Hamiltonian reduction in a
supercovariant way.
After setting supercovariant constraint,
the N=1 and N=2 super Virasoro algebras appear
in both the forms of
the Feigin-Fuchs construction and of the superschwarzian derivatives.
In this approach, it is regarded that the supersymmetric Miura
transformation is generated by a nilpotent sub-supergroup which
preserves the first-class constraints.
We also show the actions of WZW models become to the coadjoint actions
of superconformal symmetry
or those of super Liouville actions by choosing suitable
constraints, respectively. In spite of results of this paper being
not quit new, I believe the formulation of superfields may be useful
for considering properties of geometry and topology for superconformal
theories.
\par
The paper is organized as follows. Sect.2 is devoted to the
free realization of the super KM algebras and WZW models. In sect.3
 we explain the method of the Hamiltonian reduction of the
super $\rm OSp(1,2)$ KM algebra. Sect.4 is devoted to the
application for the super $\rm SL(2,1)$ KM algebra. Discussions are
contained in sect.5.
\vskip 0.5 truecm
\section{ 2. Wakimoto Construction of Super KM Algebras}
\noindent
We begin with the free field realization of KM algebras based on
Lie superalgebras discussed by Ito[11]. A root system of a Lie
superalgebra $g$ is expressed as the sum of the set of even roots
$\Delta^0$ and that of odd roots $\Delta^1$. A group element
$ g$ may be
decomposed into the direct sum of $g_0$ which is generated by the
Cartan part and the even roots and $g_1$ spanned by the odd roots.
A Kac-Moody algebra based on a Lie superalgebra of rank r is
generated by the fermionic currents $j_\alpha(x)(\alpha\in
\Delta^1)$, the bosonic currents $J_\alpha(x)(\alpha\in\Delta^0)$
and $H^i(x)(i=1,\cdot\cdot\cdot,r)$ corresponding to the Cartan
part.

In this section we extend these KM algebras to the super KM
 algebras, which have supersymmetry of two-dimensional spacetime
We concretely construct the super $\rm OSp(1,2)$ KM algebra[14].
For the first step, we construct the lagrangian of the
super ${\rm OSp}(1,2)$ WZW model.

In general, the action of the supersymmetric $\rm OSp(1,2)$ WZW model is
given by the following form:
$$
S={k \over 16\pi}\biggl\{ \int dx_+dx_-<\curm, {\cur+}>
$$
$$
- \int dx_+dx_-d\tau<g^{-1}\partial_{\tau} g,[\cur+,{\curm}]_+>\biggr\}.
\eqno(2.1)
$$
Here $x_+$ and $x_-$ are light cone coordinates and $g\coor$ is a group
element of the super $\rm OSp(1,2)$ and $<A,B>$ denotes supertrace of
 supermatrices $A$ and $B$, and $[,]_+$ indicates an anticommutation
relation.
Further, $\coder+$ and $\coderm$
\footnote"$^{1)}$"{\eightrm This notation is convenient
for actual calculations,
because we can treat as in the bosonic case.
 The chiralities of {$\scriptstyle \theta_+$ and
$\scriptstyle \theta_-$} are opposite to
an ordinary definition, but there is no problem in our discussions. }
 are covariant derivatives accompanied with $d\theta_+$ and
$d\theta_-$, respectively:
$$
\coder+ = d\theta_+D_+ = d\theta_+\biggl( {\partial \over \partial \theta_+} +
\theta_+ {\partial \over \partial x_+} \biggr),
\qquad
\coderm = d\theta_-D_- = d\theta_-\biggl( {\partial \over \partial \theta_-} +
\theta_- {\partial \over \partial x_-} \biggr).
\eqno(2.2)
$$
Now we are going to express the model in terms of free fields
as the same way as in the case of the $\rm SL(2)$ WZW model[15]. The Gauss
expansion
of a group element $g$ is given by
$$
g\coor=g_1g_2g_3=
\left(
\matrix{1      & 0      & 0                            \cr
        C      & 1      & -\Psi                         \cr
        \Psi   & 0      & 1                             \cr}
\right)
\left(
\matrix{e^{\Phi}      & 0          & 0             \cr
            0         & e^{-\Phi}  & 0              \cr
            0         & 0          & 1             \cr}
\right)
\left(
\matrix{1      & F       & \xi                       \cr
        0      & 1       & 0                        \cr
        0      & \xi     & 1                        \cr}
\right),
\eqno(2.3)
$$
\noindent
where $\Phi, C$ and $F$ are grassmann even scalar superfields and $\Psi$ and
$\xi$
are grassmann odd ones[11]. Using the Polyakov-Wiegmann relation,
the action (2.1) is rewritten in terms of the fields
$\Phi, C,F,\Psi$ and $\xi$ as
$$
S\sim\int dx_+dx_-\biggl\{-2k \hat D_-\Phi \hat D_+\Phi+e^{2\Phi}(\hat
D_-+\Psi\hat D_-\Psi)
(\hat D_+F+\xi\hat D_+\xi)-2e^\Phi \hat D_-\Psi \hat D_+\xi\biggr\}
\eqno(2.4)
$$
$$
=\int dx_+dx_-d\theta_+d\theta_-\{2kD_-\Phi D_+\Phi+BD_-C+2\Psi^\dagger
D_-\Psi\},
$$
where we define fields $B$ and $\Psi^\dagger$ as
$$
B=ke^{2\Phi}(D_+F-\xi D_+\xi),
$$
$$
\Psi^\dagger=ke^\Phi D_+\xi+{1 \over 2}B\Psi.
\eqno(2.5)
$$
This form tells the field $\Phi$ is a free scalar superfield
and the fields $(B,C)$
and ($\Psi^{\dagger},\Psi$) are superghosts. The equations of
motion are derived from invariance of the action under an infinitesimal
variation $g \rightarrow (1+\epsilon)g$:
$$
D_+(\curn)=0, \hskip 1 truecm D_-(D_+gg^{-1})=0.
\eqno(2.6)
$$
This equations indicate conservation of left currents ${\bf J}_+=kD_+gg^{-1}$
and right ones ${\bf J}_-=kg^{-1}D_-g$. Substituting $g$ into the currents
${\bf J}_+$,
we obtain the Wakimoto currents in terms of the fields
$(\Phi,C,B,\Psi,\psid)$:\footnote"$^{2)}$"{\eightrm We drop the sign + which
indicates a left
component. If signs of left and right components are dropped in equations, they
are regarded as left components.}
$$
{\bf J}=k\curr=
\left(
\matrix{H      & J^-       & j^-                            \cr
        J^+      & -H      & -j^+                         \cr
        j^+   & j^-      & 0                             \cr}
\right),
\eqno (2.7)
$$
where
$$
\left\{
\eqalign{
&j^-=\Psi^{\dagger}+{1 \over 2}B\Psi,                        \cr
&j^+=-\Psi^\dagger C+{1 \over 2}
\Psi CB-k\Psi D\Phi+kD\Psi,       \cr
&H=-\Psi\Psi^{\dagger}-CB+kD\Phi, \cr
&J^-=B,                          \cr
&J^+=-BC^2+2kCD\Phi+kDC-2\Psi\Psi^\dagger C+kD\Psi\Psi.                    \cr}
\right.
$$
It is reasonable to set the following Poisson brackets among the fields
$(\Phi,C,B,\Psi,\Psi^\dagger)$ from the action(2.4):
$$
\left\{
\eqalign{
&\{D_1\Phi_1,\Phi_2\}={1 \over 2k}\theta_{12}\delta(x_{12}),        \cr
&\{B_1,C_2\}=-\theta_{12}\delta(x_{12}),                             \cr
&\{\Psi_1^\dagger,\Psi_2\}={1 \over 2}\theta_{12}\delta(x_{12}).
\cr}
\right.
\eqno(2.8)
$$
where
$$
 \theta_{12}=\theta_1-\theta_2,\hskip 1 truecm
x_{12}=x_1-x_2-\theta_1\theta_2.
$$
{}From the definition of the currents, $H,J^+$ and $J^-$ are grassmann
odd currents which generate the super $\rm SL(2)$ Kac-Moody
algebra. $j^+$ and $j^-$ are grassmann even ones.
The classical Kac-Moody algebras can be obtained from the Poisson
brackets(2.8):
$$
\left\{
\eqalign{
\{&H_1,H_2\}={k \over 2}\delta(x_{12}),                      \cr
\{&H_1,J^{\pm}_2\}=\pm \theta_{12}J^{\pm}_2\delta(x_{12}),
\cr
\{&J^+_1,J^-_2\}=2\theta_{12}H_2\delta(x_{12})-k\delta(x_{12}),           \cr
\{&j^+_1,j^-_2\}=-\theta_{12}H_2\delta(x_{12})+{k \over 2}\delta(x_{12}),
     \cr
\{&H_1,j^{\pm}_2\}=\pm{1 \over 2} \theta_{12}j^{\pm}_2\delta(x_{12}),
\cr
\{&j^{\pm}_1,j^{\pm}_2\}=\mp{1 \over 2} \theta_{12}J^{\pm}_2\delta(x_{12}),
           \cr
\{&J^{\pm}_1,j^{\mp}_2\}=\theta_{12}j^{\pm}_2\delta(x_{12}).
\cr
}\right.
\eqno(2.9)
$$

In order to obtain the currents of quantum version, one has to consider
the quantum effect which derived from reparametrization of the measure of
a functional integral, namely, changing the fields $(\Phi,C,\Psi,F,\xi)$
to ones $(\Phi,B,C,\Psi,\Psi^{\dagger})$. This quantum effect shifts
the normalization of the scalar fields in the non-super case.
Explicit expressions of the $\rm OSp(1,2)$ currents
are written in the papers[14].
On the other hand, this shift does not appear in the super case because
 there exists two-dimensional spacetime supersymmetry.
Then there is no change in the currents of quantum version.
Canonical quantization can be done by setting the following operator
product expansions:
$$
\left\{
\eqalign{
&D_1\Phi(x_1,\theta_1)\Phi(x_2,\theta_2) \sim {1 \over 2k}{\theta_{12}
\over z_{12}},      \cr
&B(x_1,\theta_1)C(x_2,\theta_2) \sim -{\theta_{12} \over z_{12}},  \cr
&\Psi^\dagger (x_1,\theta_1)\Psi(x_2,\theta_2) \sim -{\theta_{12} \over
2z_{12}}. \cr
}\right.
\eqno(2.10)
$$

\section{ 3. Hamiltonian Reduction of Super OSp(1,2) KM Algebra}
\noindent
Now we are in a position to apply the Drinfeld and Sokolov reduction to
the super $\rm OSp(1,2)$ KM algebra realized in terms of free fields.
Bershadsky and Ooguri showed the N=1 super Virasoro algebra is derived
from the bosonic $\rm OSp(1,2)$ KM algebra by introducing
 a free fermion in addition to free fields of the Wakimoto currents[11].

Kuramoto first indicated that starting from the manifestly super
$\rm OSp(1,2)$ KM algebra, the N=1 super Virasoro algebra can be
derived without a additional fermion[14]. His approach, however, is not
explicitly supercovariant and not probably suitable for extended conformal
symmetries.

In this section we will formulate the Hamiltonian reduction of the
super $\rm OSp(1,2)$ KM algebra in a super covariant way.
Let us begin with the following first-class constraints.
$$
J^-\coor =0, \hskip 1 truecm
j^-\coor =1.
\eqno(3.1)
$$
These constraints preserve two-dimensional spacetime supersymmetry.
The components of these constraints coincide with those in the paper[14].
The form of the residual gauge transformation is written as
$$
\Omega=
\left(
\matrix{1      & 0      & 0                            \cr
        A      & 1      & -\beta                         \cr
        \beta   & 0      & 1                             \cr}
\right),
\eqno(3.2)
$$
which corresponds to a Borel sub-supergroup. Under
this transformation the differential operator ${\cal L}=kD-J$ is transformed as
$$
{\cal L} \longrightarrow \tilde \Omega {\cal L} \Omega^{-1},
\eqno(3.3)
$$
where $\tilde \Omega$ indicates changing signs of the odd fields of $\Omega$.
There exist two kinds of gauge choices, in which the phase space of
the transformed currents commute with the currents $J^-$ and
$j^-$\footnote"$^{3)}$"{\eightrm This phase space is nothing but a
quotient space divided by the residual symmetries.}. One of
these choices is called the Drinfeld and Sokolov gauge given by
$$
\Omega_1=
\left(
\matrix{1         & 0      & 0                            \cr
        j^+-kH     & 1      & -H                         \cr
        H        & 0      & 1                             \cr}
\right).
\eqno(3.4)
$$
$\difo$ is transformed as
$$
\difo_1=\tilde \Omega_1 \difo \Omega_1^{-1}
=kD-
\left(
\matrix{0      & 0      & 1                            \cr
        T     &0	     & 0                        \cr
        0        & 1      & 0                             \cr}
\right),
\eqno(3.5)
$$
where
$$
T=J^++2j^+H+kHDH+kD(j^++kDH).
\eqno(3.6)
$$
Substituting the Wakimoto currents constrained with eq.(3.1) into T,
it is rewritten as
$$
T=k^3(D\Phi\partial\Phi+\partial D\Phi).
\eqno(3.7)
$$
This form is nothing but the Feigin-Fuchs construction of the N=1
Virasoro algebra. Because of the constraint(3.1),
$$
ke^{2\Phi}D\xi =1,
\eqno(3.8)
$$
T is also expressed by the superschwarzian derivative of $\xi$:
$$
T=-k^3\biggl({\partial^2\xi \over D\xi}-2{\partial \xi(\partial D\xi)
\over (D\xi)^2}\biggr).
\eqno(3.9)
$$
We can obtain the following the classical N=1 Virasoro algebra from the Poisson
bracket
of the field $\Phi$:
$$
\{T_1,T_2\}=+{k^2 \over 2}\delta^{\prime\prime}(x_{12})
-{3 \over 2}\theta_{12}T_2\delta^\prime(x_{12})
+{1 \over 2}D_2T_2\delta(x_{12})+\theta_{12}T_2\delta(x_{12}).
\eqno(3.10)
$$

The other gauge choice is a diagonal gauge given by
$$
\Omega_2=
\left(
\matrix{1      & 0      & 0                            \cr
       -C      & 1      & \Psi                         \cr
        -\Psi   & 0      & 1                             \cr}
\right).
\eqno(3.11)
$$
Under this gauge $g$ is transformed as
$$
g \longrightarrow \Omega_2 g=
\left(
\matrix{e^{\Phi}      & 0          & 0             \cr
            0         & e^{-\Phi}  & 0              \cr
            0         & 0          & 1             \cr}
\right)
\left(
\matrix{1      & F       & \xi                       \cr
        0      & 1       & 0                        \cr
        0      & \xi     & 1                        \cr}
\right).
\eqno(3.12)
$$
Then we have the transformed currents by setting the fields
C and $\Psi$ equal zero in the currents(2.7):
$$
\difo_2=\tilde \Omega_2 \difo \Omega_2^{-1}
=kD-
\left(
\matrix{D\Phi      & 0      & 0                            \cr
        0     &-D\Phi	     & 0                        \cr
        0        & 0      & 0                             \cr}
\right).
\eqno(3.13)
$$
The phase space of this current is one of the super $\rm U(1)$
KM algebra. It is the super Miura transformation to connect the field T
with the $\rm U(1)$ KM algebra. We can express the Miura map
 in terms of the gauge transformation of the Borel sub-supergroup.
$$
\difo_1=(\tilde \Omega_1 \tilde \Omega_2^{-1})\difo_2\
( \Omega_1\Omega_2^{-1})^{-1}.
\eqno(3.14)
$$
The action (2.4) becomes a coadjoint action of the N=1
super Virasoro group by imposing the first-class constraints(3.1)
with the diagonal gauge as the second constraint:
$$
S_{super\, Virasoro} \sim \int dx_+dx_-d\theta_+d\theta_-
D_-\xi{D_+\partial_+\xi \over (D_+\xi)^2}.
\eqno(3.15)
$$
This is the same form as a geometric action for the theory
of the coadjoint orbits of the superconformal group[16][17] and its
components correspond with the result in the paper[11][13].
Imposing the similar first-class constraints for the right-handed parts
as same as eq.(3.1), one can obtain the N=1 super Liouville action
from the WZW action:
$$
S_{super\, Liouville} \sim \int dx_+dx_-d\theta_+d\theta_-
\biggl(2D_-\Phi D_+\Phi +ke^{-\Phi}\biggr).
\eqno(3.16)
$$
Supersymmetry is inherited from the structure of the Lie superalgebra
in the case of the bosonic $\rm OSp(1,2)$ KM algebra. In our case,
 starting from the manifestly supersymmetric KM algebra and
imposing the supercovariant constraints, the N=1 supersymmetry is
derived from two-dimensional spacetime.

\vskip .5 cm

\section{ 4. Hamiltonian Reduction of Super SL(2,1) KM Algebra}
\noindent
It has been shown that the Hamiltonian reduction of WZW model
based on the Lie superalgebras $\rm SL(n+1,n)(n\geq 1)$ gives the
models which have W algebra structures with the N=2 superconformal
symmetry. It is also understood that these model are the N=2 coset
models $\rm CP_n=SU(N+1)/SU(N)\times U(1)$ constructed by kazama
and Suzuki.
In this section we will discuss the simplest case of the super
$\rm SL(2,1)$ WZW model. We show that N=2 superconformal symmetry
can be obtained by imposing the constraints which do not break
N=1 supersymmetry.

The Gauss decomposition of a group element is written as follows:
$$
g\coor=
\left(
\matrix{1      & 0      & 0                            \cr
        C      & 1      & -\Psi_2                         \cr
        \Psi_1   & 0      & 1                             \cr}
\right)
\left(
\matrix{e^{\Phi_1}      & 0          & 0             \cr
            0         & e^{-\Phi_2}  & 0              \cr
            0         & 0          &  e^{\Phi_1-\Phi_2}             \cr}
\right)
\left(
\matrix{1      & F       & \xi_2                       \cr
        0      & 1       & 0                        \cr
        0      & \xi_1     & 1                        \cr}
\right).
\eqno(4.1)
$$
where $\Phi_1,\Phi_2,C$ and $F$ are grassmann even  superfields
and $\Psi_1,\Psi_2,\xi_1$ and $\xi_2$ are grassmann odd ones.
The Wakimoto currents are also given by
$$
{\bf J}=k\curr=
\left(
\matrix{H_1      & J^-       & j^-_2                            \cr
        J^+      & -H_2      & -j^+_2                           \cr
        j^+_1   & j^-_1      & H_1-H_2                          \cr}
\right).
\eqno(4.2)
$$
where
$$
\left\{
\eqalign{
H_1&=-BC-\Psi_1^\dagger\Psi_1+kD\Phi_1,                \cr
H_2&=-BC-\Psi_2^\dagger\Psi_2+kD\Phi_2,               \cr
J^-&=B,                                                 \cr
J^+&=-BC^2-C(\Psi^\dagger_1\Psi_1+\Psi^\dagger_2\Psi_2)
+kDC,                                                   \cr
&+kC(D\Phi_1+D\Phi_2)+kD\Phi_1\Psi_2\Psi_1+kD\Psi_2\Psi_1,  \cr
j^-_1&=\Psi^\dagger_1,                                   \cr
j^-_2&=\Psi^\dagger_2+B\Psi_1,                           \cr
j^+_1&=\Psi_1BC-\Psi^\dagger_2C-\Psi^\dagger_2\Psi_2\Psi_1
+kD\Phi_2\Psi_1+kD\Psi_1,                                 \cr
j^+_2&=-C\Psi^\dagger_1-k\Psi_2D\Phi_1+kD\Psi_2.          \cr}
\right.
\eqno(4.3)
$$
Here $B,\Psi_1^\dagger$ and $\Psi_2^\dagger$ are defined as
$$
B=ke^{\Phi_1}e^{\Phi_2}(DF-D\xi_1\xi_2),\quad
\Psi_1^\dagger=ke^{\Phi_2}D\xi_1+B\Psi_2,\quad
\Psi_2^
\dagger=ke^{\Phi_1}D\xi_2.
\eqno(4.4)
$$
The action is written as
\vskip 3 truecm
$$
S\sim \int dx_+dx_- d\theta_+d\theta_-\biggl(kD_-\Phi_1D_+\Phi_2
+ kD_-\Phi_2D_+\Phi_1
$$
$$
+BD_-C+D_-\Psi_1\Psi_1^\dagger+
D_-\Psi_2\Psi_2^\dagger\biggr).
\eqno(4.5)
$$
{}From this form it is natural to give the following Possion brackets
for the fields $\Phi_i,(B,C)$ and $(\Psi_i^\dagger,\Psi_i)$:
$$
\left\{
\eqalign{
&\{D\Phi_{i,1},\Phi_{j,2}\}={1 \over k}\eta_{ij}\theta_{12}\delta(x_{12}),
\cr
&\{\Psi_{i,1}^\dagger,\Psi_{j,2}\}=
\delta_{ij}\theta_{12}\delta(x_{12}), \cr
&\{B_1,C_2\}=-\theta_{12}\delta(x_{12}).   \cr}
\right.
\eqno(4.6)
$$
where $\eta_{ij}$ are elements of a matrix of which off-diagonal elements
equal one and the others equal zero.
The classical Kac-Moody algebra can be calculated as
$$
\left\{
\eqalign{
&\{H_{i,1},H_{j,2}\}={1 \over k}\eta_{ij}\delta(x_{12}),  \cr
&\{J^+_1,J^-_2\}=\theta_{12}(H_1+H_2)_2\delta(x_{12})-k\delta(x_{12}), \cr
&\{H_{i,1},j^-_{j,2}\}=-\delta_{ij}\theta_{12}j^-_{j,2}
\delta(x_{12}), \cr
&\{H_{i,1},j^+_{j,2}\}=\delta_{ij}\theta_{12}j^+_{j,2}
\delta(x_{12}), \cr
&\{j^+_{i,1},j^-_{i,2}\}=-\eta_{ij}\theta_{12}H_{j,2}\delta(x_{12})
+k\delta(x_{12}).    \cr}
\right.
\eqno(4.7)
$$
The other Poisson brackets have the similar form as the case
of $\rm OSp(1,2)$.

Let us perform the Hamiltonian reduction. First-class constraints are given by
$$
J^-\coor =0, \>
j^-_1\coor =1, \> j^-_2\coor=1.
\eqno(4.8)
$$
It is a nilpotent sub-supergroup that preserves the constraints:
$$
\Omega=
\left(
\matrix{1      & 0      & 0                            \cr
        A      & 1      & \alpha                       \cr
        -\beta   & 0      & 1                             \cr}
\right),
\eqno(4.9)
$$
Here we choose the Drinfeld and Sokolov gauge:
$$
\Omega_1=
\left(
\matrix{1         & 0      & 0                            \cr
        {1 \over 2}\{j^+_1-j^+_2-H_1H_2+kD(H_1+H_2)\}    & 1      & -H_2
\cr
        H_1        & 0      & 1                             \cr}
\right).
\eqno(4.10)
$$
In this gauge the differential operator ${\cal L}$ transforms to the form:
$$
{\cal L} \longrightarrow {\cal L_1} =
\tilde \Omega_1 {\cal L} \Omega^{-1}_1=
\left(
\matrix{0      & 0      & 1                            \cr
        T     &0	     & I                        \cr
        I        & 1      & 0                             \cr}
\right),
\eqno(4.11)
$$
where
$$
\left\{
\eqalign{
&I={1 \over 2}\{j^-_1-j^+_2-H_1H_2+kD(H_1-H_2)\},     \cr
&T=J^++H_2j^+_1+H_1j^+_2+{k \over 2}D\{j^+_1-j^+_2-H_1H_2+kD(H_1+H_2)\}
+kDH_2H_1.                                              \cr}
\right.
\eqno(4.12)
$$
Substituting the Wakimoto currents with the constrains (4.8), one can
obtain the Feigin-Fuchs construction of the N=2 super Virasoro algebra
$$
\left\{
\eqalign{
&T={k^3 \over 2}\{\partial\Phi_1D\Phi_2+\partial\Phi_2D\Phi_1
+\partial D(\Phi_1+\Phi_2)\},    \cr
&I={k^2 \over 2}\{D\Phi_1D\Phi_2+\partial\Phi_1+\partial\Phi_2\}.  \cr}
\right.
\eqno(4.13)
$$
The fields $T$ and $I$ satisfy the following the N=2 super Virasoro
algebra:
$$
\left\{
\eqalign{
&\{T_1,T_2\}=+{k^2 \over 2}\delta^{\prime\prime}(x_{12})
-{3 \over 2}\theta_{12}T_2\delta^\prime(x_{12})
+{1 \over 2}D_2T_2\delta(x_{12})+\theta_{12}T_2\delta(x_{12}), \cr
&\{T_1,I_2\}=-\theta_{12}I_2\delta^\prime(x_{12})
+{1 \over 2}D_2I_2\delta(x_{12})+\theta_{12}I_2\delta(x_{12}), \cr
&\{I_1,I_2\}={k^3 \over 2}\delta^\prime(x_{12})
-{1 \over 2}\theta_{12}T_2\delta(x_{12}) \cr}
\right.
\eqno(4.14)
$$
{}From the constraints(4.8)
$$
ke^{\Phi_1}D\xi_2=ke^{\Phi_2}D\xi_1=1.
\eqno(4.15)
$$
$T$ and $I$ are also expressed by  superschwarzian derivatives of
the fields $\xi_1$ and $\xi_2$:
$$
\left\{
\eqalign{
&T={k^3 \over 2}\biggl({\partial \xi_1D\partial \xi_2 \over D\xi_1D\xi_2}
+{\partial \xi_2D\partial \xi_1 \over D\xi_1D\xi_2}
+{\partial \xi_1D\partial \xi_1 \over D\xi_1D\xi_1}
+{\partial \xi_2D\partial \xi_2 \over D\xi_2D\xi_2}
+{\partial^2\xi_1 \over D\xi_1}
+{\partial^2\xi_2 \over D\xi_2}
\biggr),        \cr
&I={k^2 \over 2}\biggl({\partial \xi_1\partial \xi_2 \over D\xi_1D\xi_2}
+{D\partial \xi_1 \over D\xi_1}-{D\partial \xi_2 \over D\xi_2}\biggr)  \cr}
\right.
\eqno(4.16)
$$
{}From the form of (4.11) the N=2 supersymmetry comes from the structure of
the Lie superalgebra and one of two-dimensional spacetime supersymmetry.
We can also obtain the coadjoint action of the
N=2 super conformal group by taking the
first-class constraints(4.8) with the diagonal gauge:
$$
S_{super\, Virasoro}\sim\int dx_+dx_-d\theta_+d\theta_-
\biggl[\biggl\{{\partial_+D_+\xi_2 \over D_+\xi_1D_+\xi_2}
-{\partial_+\xi_2D_+\xi_2\partial_+\xi_1 \over (D_+\xi_1D_+\xi_2)^2}\biggr\}
D_-\xi_1
\eqno(4.17)
$$
$$
+\biggl\{{\partial_+D_+\xi_1 \over D_+\xi_1D_+\xi_2}
-{\partial_+\xi_1D_+\xi_1\partial_+\xi_2 \over (D_+\xi_1D_+\xi_2)^2}\biggr\}
D_-\xi_2 \biggr].
$$
As the same way as in the case of $\rm OSp(1,2)$, the N=2 super
Liouville action is obtained as
$$
S_{super\, Liouville}\sim\int dx_+dx_-d\theta_+d\theta_-
\biggl(D_-\Phi_1D_+\Phi_2+D_-\Phi_2D_+\Phi_1+e^{-\Phi_1}+e^{-\Phi_2}
\biggr).
\eqno(4.18)
$$
\section{ 5. Discussions}
\noindent
We have presented an explicit construction of the super $\rm OSp(1,2)$
and $\rm SL(2,1)$ KM algebras in terms of scalar superfields
and superghosts systems.  We could show the Feigin-Fuchs representations
and the coadjoint actions of the N=1 and N=2 super conformal symmetries
by restricting the phase space with constraints which preserve
two-dimensional spacetime supersymmetry.
In general, super-Toda theories can be obtained by use of special
Lie superalgebras with a completely fermionic simple root system[13][18].
We may obtain super W algebras and coadjoint actions of them by imposing
the constraints such as currents of simple roots equal constant.
Ito has already study the quantum Hamiltonian reduction of
super $\rm SL(n+1,n)$$(n\geq 1)$ algebras by means of an operator
formulation. Our approach needs a path integral quantization
and it is necessary to realize these algebras as gauged super
WZW models. We have to
calculate determinants derived from
gauge fixing conditions. These are under investigation. The N=2 superconformal
symmetry is especially important in both the points of studying
the compactification of superstrings and the connection with topological
field theories. So it is interesting to apply our formulation
to these theories[19].
\vskip 3mm
\endpage
\references
\vskip 3mm

\refa
\refb
\refc
\refd
\refe
\reff
\refg
\refh
\refi
\refj
\refk
\refl
\refm
\refmm
\refn
\refo
\refp
\refq
\refr
\refrr
\refs
\end